\documentclass[journal,10pt]{IEEEtran}
\usepackage{epsf}
\usepackage{graphicx}
\usepackage{subfigure}
\usepackage{amsmath}
\usepackage{theorem}
\usepackage{amssymb}
\usepackage{cite}
\usepackage{color}
\newtheorem{thm}{Theorem}

\newtheorem{lem}{Lemma}
\newtheorem{proof}{proof}

\newtheorem{rem}{Remark}

\newtheorem{DD}{Definition}
\newtheorem{EE}{Example}

\begin{document}
\title{\Huge{Error Correction for Cooperative Data Exchange}}
\author{Wentu Song,~ Xiumin Wang,~ Chau Yuen, ~Tiffany Jing Li ~and~ Rongquan Feng
\thanks{W. Song is with the School of Mathematical Sciences, Peking University, China,
        and with Singapore University of Technology and Design, Singapore.
        E-mail: songwentu@gmail.com.}
\thanks{X. Wang was with Singapore University of Technology and Design, Singapore, and with School of Computer and Information, Hefei University of Technology, Hefei 230009, China. Email: wxiumin@hfut.edu.cn.}
\thanks{C. Yuen is with Singapore University of Technology and Design,
        Singapore. Email: yuenchau@sutd.edu.sg.}
\thanks{T. J. Li is with the department of electrical and computer engineering,
        Lehigh University, Bethlehem, PA 18015, USA. Email: jingli@ece.lehigh.edu.}
\thanks{R. Feng is with the School of Mathematical Sciences,
        Peking University, China. Email: fengrq@math.pku.edu.cn.}
\thanks{This research is partly supported by the International Design Center (grant no. IDG31100102 and IDD11100101).}
}

\maketitle

\pagenumbering{arabic}

\begin{abstract}
This paper considers the problem of error correction for a
cooperative data exchange (CDE) system, where some clients are
compromised or failed and send false messages. Assuming each client
possesses a subset of the total messages, we analyze the error
correction capability when every client is allowed to broadcast only
one linearly-coded message. Our error correction capability bound
determines the maximum number of clients that can be compromised or
failed without jeopardizing the final decoding solution at each
client. We show that deterministic, feasible linear codes exist that
can achieve the derived bound. We also evaluate random linear codes,
where the coding coefficients are drawn randomly, and then develop
the probability for a client to withstand a certain number of
compromised or failed peers and successfully deduce the complete
message for any network size and any initial message distributions.
\end{abstract}
\vspace{-0.03in}
\vspace{-0.026in}

\IEEEpeerreviewmaketitle
\vspace{-0.14in}
\section{Introduction}\label{Sec.Intro}\vspace{-0.04in}
The fundamental challenge of networks is to pull together all the
available network resources and to arrange all the clients in
efficient cooperation, such that they can collaboratively deliver
a quality and trustworthy service. {\em Cooperative data exchange}
(CDE) \cite{Rouayheb09} among the clients has  become a promising
approach for achieving efficient data communications. In a CDE
system, each client initially holds only a subset of packets, and
is in quest for all the packets (from its peers). It is typical to
assume that the clients communicate through (wireless) broadcast
channels.  The objective is to design a network coded transmission
scheme that minimizes the total number of transmissions
\cite{Rouayheb09,Sprint102,Milos11} or the total transmission cost
\cite{Tajbakhsh,Ozgul11,Wang2012b}, and at the same time, ensures
all the clients can deduce the complete information.

Most of the existing studies on the CDE problem assume that the
transmission from every client is reliable and trustworthy.
However, in practice, there may exist compromised clients who
intentionally send false messages\footnote{For simplicity, we
consider error-free transmission. A transmission error may be
treated as an error-free transmission from a compromised client.},
or failed clients who send wrong readings. Such could cause
decoding error or failure, and therefore motivates us to explore
of error correction for the CDE problem. One example could be in a
sensor network, when one sensor fails, how can we detect and
correct the error through the readings of other sensors.

In the literature, several interesting works \cite{Cai,Dau1} have
studied the problem of network coding based error correction
\cite{Ahlswede,Xiumin}, but these works cannot be  applied in CDE.
This is because the existing studies assume that there exists a
single source node (sender) in possession of all the packets,
whereas in the CDE problem, there exist multiple source nodes
(senders), each equipped with only a subset of the packets.

Assuming that each client initially holds a subset of the
messages, we investigate the error correction capability that a
``fair and once'' cooperative data exchange scheme can achieve,
where ``fair and once'' means each client is allowed to broadcast
exactly one packet. We say a CDE transmission scheme is a
$\delta$-error correction solution if it guarantees the correct
recovery of the complete messages by all the clients, in the
presence of up to $\delta$ comprised clients.
 The contributions of this paper include
\begin{itemize}
\item Given initial message distribution, we derive the error
correction capability for a linear-coded CDE problem, which
specifies the maximum number of compromised clients the system can
tolerate without jeopardizing ultimate integrity and accuracy of
the message at each client. \item We show that deterministic,
feasible linear code designs exist to achieve the derived error
correction capability. \item Since deterministic coding schemes
are inflexible and unscalable, we also investigate the case of
random linear network coding. We derive the ensemble average
probability for any client to correctly deduce all the messages
despite the existence of certain compromised peers.
\end{itemize}\vspace{-0.03in}

The rest of this paper is organized as follows.
Sec.~\ref{Sec.definition} formulates the problem.
Sec.~\ref{Sec.solution} develops error correction for a general
CDE problem. We discuss the performance of random network coding
in Sec.~\ref{Sec.networkcoding}, and conclude the paper in
Sec.~\ref{Sec.conclusion}.

\vspace{-0.11in}
\section{Problem Definition and Signal Model}\label{Sec.definition}\vspace{-0.04in}
Consider a set of $k$ packets $X=\{x_1,x_2,\cdots,x_k\}$ to be
delivered to $n$ clients in $R=\{r_1,r_2,\cdots,r_n\}$, where each
message $x_i$ is assumed to be an element of a finite field
$\mathbb F$. Suppose that initially, client $r_j\in R$ holds a
subset of packets $\{x_i\}_{i\in A_j}$, and the clients
collectively have all the packets in $X$, i.e., $\bigcup_{r_j\in
R} A_j=\{1,\cdots,k\}$. To simplify the presentation, we use
$\overline{A_j}$ to denote the index set of the missing packets of
client $r_j$, i.e., $\overline{A_j}=\{1,\cdots,k\}\setminus A_j$,
and use $|\overline{A_j}|$ to denote the size of $\overline{A_j}$.
Following the system model in \cite{Rouayheb09}, the clients will
exchange packets over a common broadcast channel to assist each
other to correctly obtain all of its missing packet(s). This
problem, thereafter referred to as the cooperative data exchange
problem, is denoted by the quaternary $\mathcal H=(k,n,X,\mathcal
X)$, where $\mathcal X=\{A_1,\cdots,A_n\}$.

We assume that each client is permitted to use the common
broadcast channel \emph{exactly once}. There are $n$ clients and
each takes turn to broadcast. In the $j$th round, the client $r_j$
broadcasts an encoded packet $y_j$ which is an $\mathbb F$-linear
combination of the packets it initially has, i.e.,
$y_j=\sum_{i=1}^ka_{i,j}x_i$, where $a_{i,j}\in\mathbb F$ and
$a_{i,j}=0$ if $i\in\overline{A_j}$. The matrix
$(a_{i,j})_{k\times n}$ specifies a transmission scheme for the
CDE problem $\mathcal H=(k,n,X,\mathcal X)$ and is called an
encoding matrix of this problem.

We define the error correction problem for a general CDE problem
$\mathcal H=(k,n,X,\mathcal X)$ as follows:
\vspace{-0.12in}\begin{DD}
The {\bf $\delta$-\emph{error correction}} problem for the CDE
problem $\mathcal H=(k,n,X,\mathcal X)$
is to find a transmission scheme such that
each client $r_j$ can correctly recover all the packets in $X$,
so long as there are no more than $\delta$ compromised clients.
\end{DD}\vspace{-0.05in}
\vspace{-0.15in}
\begin{DD}\label{inci-mat-glo}
The {\bf\emph{incidence matrix}} of $\mathcal H=(k,n,X,\mathcal X)$
is defined as the matrix $C=(\xi_{i,j})_{k\times n}$, where
$\xi_{i,j}$ is a variable if $i\in A_j$, and $\xi_{i,j}=0$
otherwise. The {\bf\emph{local incidence matrix}} of $r_j$, denoted
by $C_j$, is defined as the sub-matrix of $C$, which only includes
the row vectors with indices in $\overline{A_j}$.
\end{DD}\vspace{-0.05in}
\vspace{-0.15in}
\begin{rem}\label{ecmat-inmat}
Clearly, an arbitrary encoding matrix is obtained by assigning a
value in $\mathbb F$ to each $\xi_{i,j}$ in the incidence matrix,
where $\mathbb F$ is the support field of encoding and decoding.
\end{rem}\vspace{-0.05in}
\vspace{-0.15in}
\begin{EE}\label{ex-1}:
Consider a CDE problem in which there are six messages
$x_1,\cdots,x_6$ and six clients $r_1,\cdots,r_6$, where each
message $x_i$ is an element of the ternary field $\mathbb
F_3=\{0,1,2\}$. Suppose initially, the client $r_i$ knows a subset
$A_i$ of the messages, where
$A_1=\{1,3,6\},A_2=\{2,3,4\},A_3=\{1,2,5\},A_4=\{3,4,5\},
A_5=\{2,4,6\}$ and $A_6=\{1,5,6\}$. We have the incidence matrix is
{\small\begin{equation*} C=\left(\begin{array}{cccccc}
\xi_{1,1} & 0 & \xi_{1,3} & 0 & 0 & \xi_{1,6}\\
0 & \xi_{2,2} & \xi_{2,3} & 0 & \xi_{2,5} & 0\\
\xi_{3,1} & \xi_{3,2} & 0 & \xi_{3,4} & 0 & 0\\
0 & \xi_{4,2} & 0 & \xi_{4,4} & \xi_{4,5} & 0\\
0 & 0 & \xi_{5,3} & \xi_{5,4} & 0 & \xi_{5,6}\\
\xi_{6,1} & 0 & 0 & 0 & \xi_{6,5} & \xi_{6,6}\\
\end{array}\right).\vspace{-0.05in}
\end{equation*}}\vspace{-0.03in}The local incidence
matrix of the client $r_1$ is: {\small\begin{eqnarray*}
&C_1=\left(\begin{array}{cccccc}
0 & \xi_{2,2} & \xi_{2,3} & 0 & \xi_{2,5} & 0\\
0 & \xi_{4,2} & 0 & \xi_{4,4} & \xi_{4,5} & 0\\
0 & 0 & \xi_{5,3} & \xi_{5,4} & 0 & \xi_{5,6}\\
\end{array}\right).
\end{eqnarray*}}
\end{EE}\vspace{-0.03in}


Note that the elements in $j$th column of the local incidence
matrix of client $r_j$ are all zero. This is because, the $j$th
column vector denotes the encoding vector of the packet sent by
itself, and the packets in $\{x_i\}_{i\in\overline{A_j}}$ are
unknown to $r_j$. Based on the above definition, we further define
the following matrix. \vspace{-0.12in}
\begin{DD}\label{inci-mat-glo}
Let $E=(a_{i,j})_{k\times n}$ be an \textbf{encoding matrix} of the
CDE problem $\mathcal H=(k,n,X,\mathcal X)$. The {\bf\emph{local
receiving matrix}} of $r_j$ is defined as the sub-matrix of $E$,
which includes the row vectors of $E$ with indices in
$\overline{A_j}$.
\end{DD}\vspace{-0.05in}

\vspace{-0.1in}\begin{EE}\label{ex-2} Consider the CDE problem in
Example \ref{ex-1}, the communication is completed by six rounds: in
the $i$th round, the clients $r_i$ broadcast $y_i$ to all other
clients, where $y_1=x_1+x_3+x_6$, $y_2=x_2+x_3+x_4$,
$y_3=x_1+x_2+x_5$, $y_4=x_3+2x_4+x_5$, $y_5=x_2+2x_4+x_6$ and
$y_6=x_1+2x_5+2x_6$, which is specified by the following encoding
matrix{\small\begin{equation*} E=\left(\begin{array}{cccccc}
1 & 0 & 1 & 0 & 0 & 1\\
0 & 1 & 1 & 0 & 1 & 0\\
1 & 1 & 0 & 1 & 0 & 0\\
0 & 1 & 0 & 2 & 2 & 0\\
0 & 0 & 1 & 1 & 0 & 2\\
1 & 0 & 0 & 0 & 1 & 2\\
\end{array}\right).\vspace{-0.05in}
\end{equation*}}
For $r_1$, he will receive $y_2,y_3,y_4,y_5,y_6$. Since $r_1$
knows $x_1,x_3$ and $x_6$, then he can compute
$z_2=y_2-x_3=x_2+x_4$, $z_3=y_3-x_1=x_2+x_5$,
$z_4=y_4-x_3=x_4+x_5$, $z_5=y_5-x_6=x_2+x_4$ and
$z_6=y_6-x_1-2x_6=2x_5$. So for $r_1$, he has the equation
$(0,z_2,z_3,z_4,z_5,z_6)=(x_2,x_4,x_5)E_1$ and can uniquely solve
the value of $x_2,x_4,x_5$ from $z_2,z_3,z_4,z_5,z_6$, where $E_1$
is the local receiving matrix of the client $r_1$ is
\vspace{-0.05in}{\small\begin{eqnarray*}
&E_1=\left(\begin{array}{cccccc}
0 & 1 & 1 & 0 & 1 & 0\\
0 & 1 & 0 & 2 & 2 & 0\\
0 & 0 & 1 & 1 & 0 & 2\\
\end{array}\right).
\end{eqnarray*}}
\end{EE}\vspace{-0.05in}

\vspace{-0.08in}
\section{Error Correction}\label{Sec.solution}\vspace{-0.01in}
Given the initial information held by each client, in this
section, we will first derive the error correction capability,
$\delta$, for the fair-and-once CDE problem $\mathcal
H=(k,n,X,\mathcal X)$. We will demonstrate, in the next section,
the tightness and achieveability of $\delta$ by demonstrating
feasible code designs.

To simplify the presentation, we write the packet set
$X=\{x_1,\cdots,x_k\}$ as a vector $X=(x_1,\cdots,x_k)$. For any
vector $u$, we let $\text{wt}(u)$ denote the Hamming weight of
$u$, i.e., $\text{wt}(u)$ is the number of non-zero components in
$u$. If $\mathcal C$ is a linear code of length $n$ and dimension
$k$, then $\mathcal C$ is referred to as an $[n,k]$ linear code.
Moreover, if $\mathcal C$ has a minimum distance $d$, then
$\mathcal C$ is referred to as an $[n,k,d]$ linear code.

For a CDE problem $\mathcal H=(k,n,X,\mathcal X)$, we define the
\emph{information space} of $r_j$ as follows:
\vspace{-0.12in}\begin{DD} The {\bf\em information space} of
$r_j$, denoted by $V_j$, is defined as the set of all possible
packets of $X$ estimated by the client $r_j$, i.e.,
$V_j=\{(\hat{x}_1,\cdots,\hat{x}_k)\in\mathbb F^k;
\hat{x}_i=x_i\text{~if~}i\in A_j\}.$
\end{DD}\vspace{-0.08in}

Clearly, $V_j$ contains altogether $|\mathbb
F|^{|\overline{A_j}|}$ vectors, corresponding to all the
exhaustive trial decoding solutions, yet only one is the true and
correct message solution. For example, in Example~\ref{ex-1}, the
information space of $r_1$ is
$V_1=\{(x_1,\hat{x}_2,x_3,\hat{x}_4,\hat{x}_5,x_6);
\hat{x}_2,\hat{x}_4,\hat{x}_5\in\mathbb F\}$.

Since the client $r_j$ knows packets $x_i$, $\forall i\in A_j$, it
must determine, from its received message vector
$Y=(y_1,\cdots,y_n)$, a candidate vector $\hat{X}\in V_j$ as its
decoder output. Given a received vector $Y\in \mathbb F^n$, the
\emph{minimum distance decoder} of $r_j$ is a map $
 \mathfrak D : \mathbb F^n\rightarrow V_j
$ such that the decoder output $\mathfrak D(Y)$ satisfies
$d_H(\mathfrak D(Y)E,Y)\leq d_H(\hat{X}E,Y)$ for any $\hat{X}\in
V_j$, where $d_H(\cdot,\cdot)$ is the Hamming distance function.

\vspace{-0.12in}
\begin{lem}\label{erc-con}
A transmission scheme with the encoding matrix $E$ is a
$\delta$-error correction solution of the CDE problem $\mathcal
H=(k,n,X,\mathcal X)$ if and only if each local receiving matrix
$E_j$ is a generating matrix of an $[n,|\overline{A_j}|]$ linear
code with minimum distance $d\geq 2\delta+1$.
\end{lem}
\vspace{-0.08in}
\begin{proof}
From the theory of classical error-correcting codes
\cite{MacWilliams}, for any client $r_j\in R$, the transmission
scheme can correct $\delta'\leq\delta$ errors if and only if for
any $\hat{X},\hat{X}'\in V_j$ and $\hat{X}\ne \hat{X}'$,
$d_H(\hat{X}E,\hat{X}'E)\geq 2\delta+1$, or, equivalently,
{\small\begin{eqnarray} \text{wt}(\hat{X}E-\hat{X}'E)\geq
2\delta+1, \forall\{\hat{X},\hat{X}'\}\subseteq V_j
\end{eqnarray}}
\vspace{0.03in}Let $U_j=\{(\hat{x}_1,\cdots,\hat{x}_k)\in\mathbb
F^k; \hat{x}_{i}=0, \forall i\in A_j\}\setminus \{0_k\}$, where $
0_k$ denotes all-zero row vector of length $k$. Note that
$\hat{X}E-\hat{X}'E=(\hat{X}-\hat{X}')E$. It is easy to see that
$U_j=\{\hat{X}-\hat{X}'; \hat{X},\hat{X}'\in V_j$ and $\hat{X}\ne
\hat{X}'\}$, so Eq. (1) is equivalent to
{\small\begin{eqnarray}\text{wt}(\hat{X}E)\geq 2\delta+1,
\forall\hat{X}\in U_j
\end{eqnarray}}\vspace{-0.05in}
By the definition of $E_j$ and $U_j$, Eq.~(2) is equivalent to
\vspace{-0.05in}{\small\begin{eqnarray}\text{wt}(\tilde{X}E_j)\geq
2\delta+1, \forall\tilde{X}\in\mathbb
F^{|\overline{A_j}|}\setminus \{ 0_{|\overline{A_j}|}\}
\end{eqnarray}}
\vspace{-0.01in} Eq. (3) means that $E_j$ is a generating matrix
of an $[n,|\overline{A_j}|]$ linear code with minimum weight at
least $2\delta+1$. Note that the minimum distance of a linear code
equals to the minimum weight of it, and this proves
Lemma~\ref{erc-con}.
\end{proof}\vspace{-0.01in}

Lemma \ref{erc-con} shows an important relation between the
error-correction capability of a transmission scheme and the
minimum distance of the linear code generated by the corresponding
local receiving matrices. The following lemma gives a method to
determine the minimum distance of a linear code from its
generating matrix.\vspace{-0.11in}
\begin{lem}\label{g-mat}
Suppose $\mathcal C$ is an $[n,k]$ linear code and $G$ is a
generating matrix of $\mathcal C$. Then the minimum distance of
$\mathcal C$ is at least $d$ if and only if any sub-matrix
including $n-d+1$ columns of $G$ has rank $k$ (where $n-d+1\ge
k$).
\end{lem}\vspace{-0.08in}
\begin{proof}
Let $G_1,\cdots,G_n$ be the $n$ columns of $G$. Note that the
minimum distance of a linear code equals to the minimum weight of
it. The minimum distance of $\mathcal C$ is at least $d$ if and
only if the minimum weight of $\mathcal C$ is at least $d$. That
is, for any $X\in\mathbb F^{k}\setminus\{0_k\}$,
$\text{wt}(XG)\geq d$. Clearly, this condition is equivalent to
the following condition\\ ($*$): For any $X\in\mathbb
F^{k}\setminus\{0_k\}$, the vector $XG$ has at most $n-d$ zero
elements.

For any $\{i_1,\cdots, i_{n-d+1}\}\subseteq\{1,\cdots,n\}$, consider
the system of linear equations {\small\begin{eqnarray}\label{4}
\hat{X}(G_{i_1},\cdots,G_{i_{n-d+1}})=0_{n-d+1}\end{eqnarray}}where
$\hat{X}$ is a vector of $k$ variables. Note that Eq. (\ref{4}) are
$n-d+1$ equations. Then condition ($*$) holds if and only if for any
$X\in\mathbb F^{k}\setminus\{0_k\}$, $X$ is not a solution of
Eq.~(4), which means that Eq. (\ref{4}) has only zero solution,
i.e., only $0_k$ is the solution of it. By the knowledge of linear
algebra, Eq. (\ref{4}) has only zero solution if and only if the
submatrix $(G_{i_1},\cdots,G_{i_{n-d+1}})$ has rank $k$. Thus, the
minimum distance of $\mathcal C$ is at least $d$ if and only if any
submatrix including $n-d+1$ columns of $G$ has rank $k$.
\end{proof}\vspace{-0.02in}

By remark \ref{ecmat-inmat}, designing a $\delta$-error correction
solution of a CDE problem is equivalent to assign a value in
$\mathbb F$ to each variable $\xi_{i,j}$ in the incidence matrix
such that the resulted local receiving matrices satisfy the
condition of Lemma \ref{g-mat} $($for their own parameters$)$. In
the following, we focus on the incidence matrix of the CDE problem
$\mathcal H=(k,n,X,\mathcal X)$.

We use $\mathbb F[\xi_1,\cdots,\xi_N]$ to denote the polynomial
ring of the variables $\xi_1,\cdots$, $\xi_N$ over the field
$\mathbb F$. Let $r$ be a positive integer. An $r\times r$ matrix
$L$ over the ring $\mathbb F[\xi_1,\cdots,\xi_N]$ is said to be
non-singular if the determinant of $L$ is a nonzero polynomial in
$\mathbb F[\xi_1,\cdots,\xi_N]$.

\vspace{-0.08in}
\begin{DD}\label{diam-loc}
Suppose $M$ is an $r\times l ~ (r\leq l)$ matrix over $\mathbb
F[\xi_1,\cdots,\xi_N]$. The {\bf\emph{diameter}} of $M$ is defined
as the smallest positive integer $\rho$ such that any $\rho$
columns of $M$ contain an $r\times r$ non-singular sub-matrix.
\end{DD}\vspace{-0.1in}

For a given CDE problem $\mathcal H=(k,n,X,\mathcal X)$, let
$\rho_j$ be the diameter of the local incidence matrix of $r_j,
j=1,\cdots,n$. We define the diameter of $\mathcal H$ as
$\rho=\text{max}\{\rho_1,\cdots,\rho_n\}$.

Reconsider the CDE problem in Example \ref{ex-2}. It is easy to
verify that $\rho_j=4, \forall j\in\{1,\cdots,6\}$. Thus the
diameter of $\mathcal H$ in this example is $\rho=4$.

\vspace{-0.1in}
\begin{DD}\label{char-poly}
For a CDE problem $\mathcal H=(k,n,X,\mathcal X)$, let $\mathcal
L_j=\{L; L \text{~is a non-singular square sub-matrix of~} C_j
\text{~of order~} |\overline{A_j}|\}$ and $\mathcal
L=\cup_{j=1}^n\mathcal L_j$. We then define the {\bf
\emph{character polynomial}} of $\mathcal H$ as the polynomial
\vspace{-0.02in}{\small
\begin{eqnarray*}h(\cdots,\xi_{i,j},\cdots)=\prod_{L\in\mathcal
L}\text{det}(L)\end{eqnarray*}} where $\text{det}(L)$ is the
determinant of the square matrix $L$.
\end{DD}\vspace{-0.02in}

The following lemma transfers the problem of designing a
$\delta$-error correction solution of $\mathcal H$ to the problem
of finding a nonzero point of the character polynomial of
$\mathcal H$. \vspace{-0.12in}\begin{lem}\label{poly-solu} Let
$h(\cdots,\xi_{i,j},\cdots)$ be the character polynomial of
$\mathcal H$ and $E=(a_{i,j})$ be an encoding matrix of $\mathcal
H$ such that $h(\cdots,a_{i,j},\cdots)\ne 0$. Then $E$ is a
$\lfloor\frac{n-\rho}{2}\rfloor$-error correcting solution of
$\mathcal H$, where $\rho$ is the diameter of $\mathcal H$.
\end{lem}\vspace{-0.06in}
\begin{proof}\vspace{-0.02in}
Let $C_j$ and $E_j$ be the local incidence matrix and the local
receiving matrix of client $r_j$. Since $\rho$ is the diameter of
$\mathcal H$, any set of $\rho$ columns of $C_j$ contains a
non-singular sub-matrix $L$ of order $|\overline{A_j}|$, i.e.,
$L\in\mathcal L$. Correspondingly, any set of $\rho$ columns of
$E_j$ contains a sub-matrix $L'$ such that $L'$ is obtained by
replacing $\xi_{i,j}$ by $a_{i,j}$ in $L$. Since
$h(\cdots,a_{i,j},\cdots)\ne 0$, we have $\text{det}(L')\neq 0$. As
$L'$ has rank $|\overline{A_j}|$, and it follows that any set of
$\rho$ columns of $E_j$ has rank $|\overline{A_j}|$. According to
Lemma \ref{g-mat}, $E_j$ is a generating matrix of an
$[n,|\overline{A_j}|]$ linear code and its minimum distance $d\geq
n-\rho+1$. Let $\delta=\lfloor\frac{n-\rho}{2}\rfloor$. Then, we
have $2\delta\leq n-\rho$. Thus, $2\delta+1\leq d$. According to
Lemma \ref{erc-con}, $E$ is the encoding matrix of a
$\lfloor\frac{n-\rho}{2}\rfloor$-error correcting solution of
$\mathcal H$.
\end{proof}\vspace{-0.01in}

To make further discussion, we need the following lemma, which is
a well-known result in algebra (e.g., see \cite{Koetter03}).
\vspace{-0.1in}
\begin{lem}\label{pol-zero}
Let $f(\xi_1,\cdots,\xi_N)$ be a nonzero polynomial in $\mathbb
F[\xi_1,\cdots,\xi_N]$. For a sufficiently large field $\mathbb
F$, there exists an $n$-tuple $(a_1,\cdots,a_N)\in\mathbb F^N$
such that $f(a_1,\cdots,a_N)\neq 0$.
\end{lem}\vspace{-0.03in}

Now, we can prove our main result for deterministic coding.
\vspace{-0.2in}\begin{thm}\label{eccde-main} Suppose $\mathbb F$
is sufficiently large. Then the CDE problem $\mathcal
H=(k,n,X,\mathcal X)$ has a $\delta$-error correcting solution if
and only if $\delta\leq \lfloor\frac{n-\rho}{2}\rfloor$, where
$\rho$ is the diameter of $\mathcal H$.
\end{thm}\vspace{-0.03in}
\vspace{-0.02in}\begin{proof} We first prove the sufficiency of
the condition by assuming that $\delta\leq
\lfloor\frac{n-\rho}{2}\rfloor$. According to Lemma
\ref{pol-zero}, there exists a feasible assignment for each
$\xi_{i,j}$ with $a_{i,j}\in\mathbb F$, such that
$h(\cdots,a_{i,j},\cdots)\ne 0$. By Lemma \ref{poly-solu},
$E=(a_{i,j})$ is the encoding matrix of a
$\lfloor\frac{n-\rho}{2}\rfloor$-error correcting solution of
$\mathcal H$. That is, $E$ is the encoding matrix of a
$\delta$-error correcting solution.

We then prove the necessity of the condition, where we assume that
$\mathcal H$ has a $\delta$-error correcting solution with
encoding matrix $E=(a_{i,j})_{k\times n}$. By Lemma \ref{erc-con},
each local encoding matrix $E_j$ is a generating matrix of a
$[n,|\overline{A_j}|]$ linear code with minimum distance $d\geq
2\delta+1$. By Lemma \ref{g-mat}, any set of $n-d+1$ columns of
$E_j$ has rank $|A_j|$, i.e., any set of $n-d+1$ columns of $E_j$
contains a non-singular sub-matrix of order $|A_j|$.
Correspondingly, any set of $n-d+1$ columns of $C_j$ contains a
non-singular sub-matrix of order $|A_j|$ over the ring $\mathbb
F[\cdots,\xi_{i,j},\cdots]$. We have $\rho_j\leq n-d+1$ and
$\rho=\text{max}\{\rho_1,\cdots,\rho_n\}\leq n-d+1$, where
$\rho_j$ is the diameter of the local incidence matrix $C_j$ of
$r_j$. Thus, $d\leq n-\rho+1$. Combining the afore-proven result
that $d\geq 2\delta+1$, we can deduce that $2\delta\leq n-\rho$.
Since $\delta$ is an integer, we have
$\delta\leq\lfloor\frac{n-\rho}{2}\rfloor$. Thus, we complete the
proof of Theorem \ref{eccde-main}.
\end{proof}\vspace{-0.02in}

Consider the CDE problem $\mathcal H=(6,6,X,\mathcal X)$ in
Example \ref{ex-1}, of which we have shown that the diameter is 4.
So by Theorem \ref{eccde-main}, $\mathcal H$ has a $\delta$-error
correcting solution for any $\delta\leq 1$, which means the system
can tolerate at most one compromised client; otherwise, some
clients may not be able to correctly deduce all the messages. It
can also be easily verified that the encoding strategy given by
the encoding matrix $E$ in Example \ref{ex-2} achieves the derived
capability $\delta\le 1$. In other words, if any one  client is
compromised and sends a false message intentionally, the encoding
strategy given by matrix $E$ can detect the false message and make
sure all the clients can successfully decode their missing packets
using their local receiving matrices.

For example, if among $y_2,y_3,y_4,y_5,y_6$ in Example \ref{ex-2}
there is one, say $y_2$, which change to a erroneous value $y_2'$,
then $z_2$ will also change to a erroneous value $z_2'=y_2'-x_3$.
Since $\mathcal C$ is with minimum distance $3$,
$(0,z_2,z_3,z_4,z_5,z_6)$ is the the nearest codeword to
$(0,z_2',z_3,z_4,z_5,z_6)$, i.e., for any
$(x_2',x_4',x_5')\ne(x_2,x_4,x_5)$, $(x_2',x_4',x_5')E_1$ has at
least two elements different from $(0,z_2',z_3,z_4,z_5,z_6)$. By
the minimum distance decoder, we can still obtain the correct
value of $x_2,x_4,x_5$.

\section{Performance with Random Network Coding}\label{Sec.networkcoding}\vspace{0.02in}
Although there exists feasible deterministic code designs to
realize the $\delta$ error-correction, the deterministic encoding
matrix must be defined and distribute across the network system
beforehand. This not only incurs extra communication overhead, but
also makes the system rather inadaptive and unscalable, as any
change in the network size, or in the individual packet sets
possessed by the clients, will cause a re-computation and
re-distribution of the entire coding scheme. To make the system
more robust, scalable and hence more practical, we now consider
using random linear network codes and evaluate its performance.

In the distributed, random coding context, each client locally and
independently generates an encoded packet over its possession, and
broadcasts to all of its peers. The coefficients of the encoding
vector are randomly selected from a predefined field $\mathbb F$.
Again, assuming there exist malicious clients, we are interested
in the computing the error tolerance capability of the system.
Unlike the deterministic case, here the error tolerance must be
evaluated over the ensemble of the random coding schemes, assuming
each and every instance is equally probable. The analytical result
is therefore represented in terms of the probability.

Before further analysis, we introduce the following
Schwartz-Zippel Lemma~(e.g., see \cite{Motwani}).
\vspace{-0.05in}\begin{lem}\label{sch-zip} Let
$f(\xi_1,\cdots,\xi_N)$ be a nonzero polynomial of degree $d\geq
0$ over a field $\mathbb F$. Let $S$ be a finite subset of
$\mathbb F$, and the value of each $\xi_1,\cdots,\xi_N$ be
selected independently and uniformly at random from $S$. Then the
probability that the polynomial equals zero is at most
$\frac{d}{|S|}$, i.e.,
$\Pr(f(\xi_1,\cdots,\xi_N)=0)\leq\frac{d}{|S|}$.
\end{lem}\vspace{-0.03in}

We now prove our random coding result:
\vspace{-0.1in}\begin{thm}\label{rand-code} Suppose that the
character polynomial of the CDE problem $\mathcal
H=(k,n,X,\mathcal X)$ is of degree $d$ and the size of the field
$\mathbb F$ is $q>d$. Let the encoding coefficients $\{a_{i,j}\}$
be chosen independently and uniformly at random from $\mathbb F$.
Then the probability that $E=(a_{i,j})$ is the encoding matrix of
a $\lfloor\frac{n-\rho}{2}\rfloor$-error correcting solution of
$\mathcal H$ is at least $1-\frac{d}{q}$, where $\rho$ is the
diameter of $\mathcal H$.
\end{thm}
\vspace{-0.08in}
\begin{proof}
Let $h(\cdots,\xi_{i,j},\cdots)$ be the character polynomial of
$\mathcal H$. According to Lemma \ref{sch-zip}, by randomly
selecting $a_{i,j}$ in the field $\mathbb F$,
Pr$(h(\cdots,a_{i,j},\cdots)=0)\leq \frac{d}{q}$. Hence,
Pr$(h(\cdots,a_{i,j},\cdots)\neq 0)\geq 1-\frac{d}{q}$. From Lemma
\ref{poly-solu}, the probability that $E=(a_{i,j})_{k\times n}$ is
the encoding matrix of a $\lfloor\frac{n-\rho}{2}\rfloor$-error
correcting solution of $\mathcal H$ is at least $1-\frac{d}{q}$.
\end{proof}\vspace{0.03in}

{\bf Remark 2:} Clearly, the degree of the character polynomial of
the CDE problem $\mathcal H=(k,n,X,\mathcal X)$ only depends on
the parameters $k,n$ and $\mathcal X$ and is independent of the
field $\mathbb F$. By Theorem \ref{rand-code}, if the field
$\mathbb F$ is sufficiently large, with a high probability, we can
obtain a $\lfloor\frac{n-\rho}{2}\rfloor$-error correcting
solution of $\mathcal H$ by randomly choosing the encoding
coefficients from the given field. \vspace{-0.12in}
\section{Conclusion}\label{Sec.conclusion}\vspace{-0.01in}
We have studied the error correction capability for a network coded
data exchange problem. Assuming every client in the network is
allowed to exchange only one message, we develop a tight upper bound
on the maximal clients that can be compromised or failed without
undermining the final messages. We show that deterministic schemes
exist to achieve the bound. For the system to be more scalable, we
also consider random coding, and develop the probability that each
client can successfully identify the erroneous message and deduce
the complete information. It is worthy remark that since the
encoding matrix is restrict, the construction technique in classical
linear code can not apply to the CDE error correction code. Thus, we
give rise to a new problem in code design.

\appendices

\ifCLASSOPTIONcaptionsoff
  \newpage
\fi 








\end{document}